\documentclass[12pt]{article}%
\usepackage{amssymb}
\usepackage{graphicx}
\usepackage{amsmath}%
\usepackage{amsfonts}
\newtheorem{theorem}{Theorem}

\newtheorem{conjecture}[theorem]{Conjecture}

\newtheorem{lemma}[theorem]{Lemma}

\newtheorem{proposition}[theorem]{Proposition}

\newenvironment{proof}[1][Proof]{\noindent\textbf{#1.} }{\ \rule{0.5em}{0.5em}}
\begin{document}

\title{\textbf{3D crystal: how flat its flat facets are?}}
\author{\textbf{Thierry Bodineau}\\Universit\'{e} Paris 7, D\'{e}partement de Math\'{e}matiques, \\Case 7012, 2 place Jussieu, F-75251 Paris, France\\Thierry.Bodineau@gauss.math.jussieu.fr
\and \textbf{Roberto H. Schonmann}\\Mathematics Department, UCLA, \\Los Angeles, CA 90095, U.S.A.\\rhs@math.ucla.edu
\and \textbf{Senya Shlosman}\\Centre de Physique Theorique, CNRS,\\Luminy Case 907,\\13288 Marseille, France\\and \\IITP, RAS, Moscow 101477, Russia\\shlosman@cpt.univ-mrs.fr}
\maketitle

\begin{abstract}
We investigate the hypothesis that the (random) crystal of the $\left(
-\right)  $-phase inside the $\left(  +\right)  $-phase of the 3D canonical
Ising model has flat facets. We argue that it might need to be weakened, due
to the possibility of formation of extra monolayer on a facet. We then prove
this weaker hypothesis for the Solid-On-Solid model.

\end{abstract}

\section{Introduction}

Consider the classical Ising model $\left\{  \sigma_{s}=\pm1\right\}  ,$ given
by the Hamiltonian
\[
H^{\text{Ising}}\left(  \sigma\right)  =-\sum_{\substack{s,t\in\mathbb{Z}^{d}:
\\\left\vert s-t\right\vert =1}}\sigma_{s}\sigma_{t},
\]
in the cubic box $\mathbb{T}_{N}\subset\mathbb{Z}^{d}$ of side $N,$ with
periodic boundary conditions and at the temperature $\beta^{-1},$ which is low
enough. Let us impose the canonical constraint:
\[
\sum_{s\in\mathbb{T}_{N}}\sigma_{s}=b\left\vert \mathbb{T}_{N}\right\vert ,
\]
where $\left\vert \mathbb{T}_{N}\right\vert $ is the volume of the box
$\mathbb{T}_{N},$ the constant $b$ satisfies $\left(  1-\left(  \frac{1}%
{d}\right)  ^{d/\left(  d-1\right)  }\right)  <b<m\left(  \beta\right)  ,$ and
where $m\left(  \beta\right)  $ is the spontaneous magnetization. Then the
typical configuration $\sigma$ under the Gibbs canonical distribution will
have a crystal: namely, it will have one large contour $\Gamma\left(
\sigma\right)  $ (which is a surface of codimension one, of linear size $\sim
N$ and of volume $\sim N^{d}$), randomly located, such that inside $\Gamma$ we
will see the minus-phase, while outside $\Gamma$ the plus-phase.\footnote{Of
course we will have the separation of phases for all values of $b.$ Our
restriction excludes the case when the minus-phase arranges itself into a
strip, wrapping around the torus.}

We are interested in the geometry of the (random) crystal $\Gamma\left(
\sigma\right)  .$ It is known that under a certain scaling the shape of
$\Gamma$ tends to a non-random limit. Namely, if one shifts $\Gamma$ so that
its center of mass will be at the origin, and then scales $\Gamma$ by a factor
of $\frac{1}{N}$ in every direction, then in the limit $N\rightarrow\infty$
the random surface $\frac{1}{N}\Gamma$ will approach the non-random surface
$W_{d}\left(  \beta\right)  ,$ the well-known Wulff shape. (In fact,
$W_{d}\left(  \beta\right)  $ depends also on $b,$ but since this dependence
is just a linear scaling, we will omit it.) The meaning of the word
\textquotedblleft approach\textquotedblright\ depends on the dimension $d.$ In
dimension 2 the surface $W_{2}\left(  \beta\right)  $ is just an analytic
curve, and a question of convergence of $\Gamma$ to $W_{2}\left(
\beta\right)  $ is studied in \cite{DKS,IS} in great detail. Namely, with
probability going to 1 as $N\rightarrow\infty,$ one can shift the curve
$NW_{2}\left(  \beta\right)  $ in such a way that the contour $\Gamma$ will
lie inside the $N^{3/4}$-neighborhood of $NW_{2}\left(  \beta\right)  .$ In
dimension 3 the known results hold in a weaker $L^{1}$ sense: one should pass
from the configuration $\sigma$ to its integrated magnetization profile, which
is a function $M_{\sigma}\left(  x\right)  \in\left[  -1,+1\right]  $ on the
unit torus $\mathcal{T}^{3}.$ Then the $L^{1}$ distance between $\frac
{1}{m\left(  \beta\right)  }M_{\sigma}\left(  x\right)  ,$ properly shifted,
and the signed characteristic function $2\mathbf{I}_{W_{3}\left(
\beta\right)  }-1$ of the inside of the surface $W_{3}\left(  \beta\right)  $
goes to zero as $N\rightarrow\infty,$ see \cite{CP, B, BIV}. Notice that on a
suitable coarse grained scale, refined results on the stability of the Wulff
crystal w.r.t. the Hausdorff distance were obtained in \cite{BI}.

Unlike the curve $W_{2}\left(  \beta\right)  ,$ the surface $W_{3}\left(
\beta\right)  $ is not analytic; moreover, it contains flat pieces -- called
facets -- provided that the temperature $\beta^{-1}$ is below certain critical
temperature $T_{r}$ -- called roughening temperature. It is known rigorously
that $T_{r}\geq T_{c}\left(  2\right)  ,$ see \cite{BFL,BFM},\textbf{\ }where
we denote by $T_{c}\left(  d\right)  $ the critical temperature of the
$d$-dimensional Ising model. It is an open question whether $T_{r}$ is equal
to $T_{c}\left(  3\right)  $ or is strictly less, as the common belief is. The
shape of the facets of $W_{3}\left(  \beta\right)  $ is also given by the
Wulff construction, see \cite{M} or \cite{S1}, sect. 2.5.

On the microscopic level, it was proven by Dobrushin \cite{D} that at
sufficiently low temperatures, rigid interfaces occur for some Gibbs measures
with specific choices of boundary conditions. Therefore it is a natural
question to ask, in which sense the flat facets observed in the macroscopic
crystals and the microscopic rigid interfaces are related. In this paper we
want to discuss the question of whether or not the random crystals
$\Gamma\left(  \sigma\right)  $ themselves have flat facets, for $N$ large.
Clearly, the results concerning the $L^{1}$-convergence of $\Gamma\left(
\sigma\right)  $ to $W_{3}\left(  \beta\right)  $ are perfectly consistent
with either behavior. Some time ago one of us made the following conjecture,
see \cite{S1}, sect. 3.4:

\begin{conjecture}
(\textbf{Probably wrong}) Let the temperature $\beta^{-1}\ $be low enough.
Then the following event has probability approaching $1$ as $N\rightarrow
\infty:$

\noindent There exist six distinct 2D planes $L_{i}=L_{i}\left(
\sigma\right)  \subset\mathbb{T}_{N},$ $i=1,...,6,$ two for each coordinate
direction, such that the intersections $L_{i}\cap\Gamma\left(  \sigma\right)
$ are flat facets of $\Gamma\left(  \sigma\right)  .$ Namely, for every $i$

\noindent$i)$ $\,\mathrm{diam}\left(  L_{i}\cap\Gamma\left(  \sigma\right)
\right)  \geq C_{1}\left(  \beta\right)  \mathrm{diam}\left(  \Gamma\left(
\sigma\right)  \right)  $, with $C_{1}\left(  \beta\right)  \rightarrow$%
$\sqrt{2/3}$ as $\beta\rightarrow\infty;$

\noindent$ii)$ $\,\frac{\mathrm{Area}\left(  L_{i}\cap\Gamma\left(
\sigma\right)  \right)  }{\left[  \mathrm{diam}\left(  L_{i}\cap\Gamma\left(
\sigma\right)  \right)  \right]  ^{2}}\geq C_{2}\left(  \beta\right)  ,$ with
$C_{2}\left(  \beta\right)  \rightarrow1/2$ as $\beta\rightarrow\infty,$ where by

\noindent$\mathrm{Area}\left(  L_{i}\cap\Gamma\left(  \sigma\right)  \right)
$ we mean the number of plaquettes of $\Gamma\left(  \sigma\right)  , $
belonging to the plane $L_{i};$

\noindent$iii)$ \thinspace The asymptotic shape of the facets $L_{i}\cap
\Gamma\left(  \sigma\right)  $ is given by the corresponding Wulff
construction, see \cite{M} or \cite{S1}, sect. 2.5.
\end{conjecture}

We believe now that the above statement is a little bit too strong to be true.
More precisely, it is almost true, except that one of the above 6 facets has
an extra monolayer of the height one! So our refined conjecture looks as follows:

\begin{conjecture}
(\textbf{Hopefully correct}) Let the temperature $\beta^{-1}\ $be low enough.
Then the following event has probability approaching $1$ as $N\rightarrow
\infty:$

\noindent There exist six distinct 2D planes $L_{i}=L_{i}\left(
\sigma\right)  \subset\mathbb{T}_{N},$ $i=1,...,6,$ two for each coordinate
direction, such that the intersections $L_{i}\cap\Gamma\left(  \sigma\right)
$ are flat facets of $\Gamma\left(  \sigma\right)  $ in the following sense:

\begin{itemize}
\item $\,$for every $i$ $\mathrm{diam}\left(  L_{i}\cap\Gamma\left(
\sigma\right)  \right)  \geq C_{1}\left(  \beta\right)  \mathrm{diam}\left(
\Gamma\left(  \sigma\right)  \right)  $, with $C_{1}\left(  \beta\right)
\rightarrow$$\sqrt{2/3}$ as $\beta\rightarrow\infty;$

\item for every $i$ except $i=i_{0}=i_{0}\left(  \sigma\right)  $ $\,$

$\frac{\mathrm{Area}\left(  L_{i}\cap\Gamma\left(  \sigma\right)  \right)
}{\left[  \mathrm{diam}\left(  L_{i}\cap\Gamma\left(  \sigma\right)  \right)
\right]  ^{2}}\geq C_{2}\left(  \beta\right)  ,$ with $C_{2}\left(
\beta\right)  \rightarrow1/2$ as $\beta\rightarrow\infty;$

\item $\frac{\mathrm{Area}\left(  L_{i_{0}}\cap\Gamma\left(  \sigma\right)
\right)  +\mathrm{Area}\left(  \left(  L_{i_{0}}+\mathbf{n}_{i_{0}}\right)
\cap\Gamma\left(  \sigma\right)  \right)  }{\left[  \mathrm{diam}\left(
L_{i_{0}}\cap\Gamma\left(  \sigma\right)  \right)  \right]  ^{2}}\geq
C_{2}\left(  \beta\right)  ,$ where $\mathbf{n}_{i_{0}}$ is the unit vector
orthogonal to $L_{i_{0}}$ and pointing \textquotedblleft away
from\textquotedblright\ $\Gamma\left(  \sigma\right)  .$
\end{itemize}
\end{conjecture}

The meaning of the last statement is that on the facet $L_{i_{0}}\cap
\Gamma\left(  \sigma\right)  $ there is another \textquotedblleft
monoatomic\textquotedblright\ layer of our crystal, having the shape $\left(
L_{i_{0}}+\mathbf{n}_{i_{0}}\right)  \cap\Gamma\left(  \sigma\right)  .$ The
limiting values $\sqrt{2/3}$ and $1/2$ are coming from the fact that in the
limit $\beta\rightarrow\infty$ we expect $\Gamma\left(  \sigma\right)  $ to
approach the shape of the cube.

At present we have no proof of this conjecture, and our paper is a result of
an attempt to prove it. Namely, we prove here a weaker statement, and for a
simpler -- SOS -- model. More precisely, we show that in the \textquotedblleft
canonical\textquotedblright\ SOS-model indeed a flat facet is formed, which
may have an extra monolayer of particles. We formulate our result in the next
section. In section 3 we further discuss it and we make various comments
concerning the validity of the conjectures above.\textbf{\ }

The heuristic explanation of our result is simple. Imagine that on one facet
of the crystal $\Gamma\left(  \sigma\right)  $ we have two monolayers -- the
top one, $F_{1}\left(  \sigma\right)  ,$ located over the second one,
$F_{2}\left(  \sigma\right)  ,$ with the size of the second one significantly
smaller than the size of $\Gamma\left(  \sigma\right)  $ itself. Then we can
enlarge $F_{2}\left(  \sigma\right)  $ to the full size of the facet of
$\Gamma\left(  \sigma\right)  ,$ diminishing at the same time the monolayer
$F_{1}\left(  \sigma\right)  .$ It might even be that by that procedure the
monolayer $F_{1}\left(  \sigma\right)  $ will disappear completely. But in any
case this procedure decreases the surface energy of the crystal. The reason
for that is the same as for the fact that merging together two droplets into a
larger one decreases the overall surface energy. In fact, we need here a
slightly more general statement, that if the possible growth of the larger
droplet is constrained by the container, then still to grow it to the maximal
possible size, while diminishing the smaller one correspondingly is
energetically favorable. In the present paper it will be enough for us to have
a zero-temperature analog of this statement, which is the content of the Lemma
\ref{drop} below. This statement for the general case can be proven by the
methods of the paper \cite{SS1}.

The second result of our paper deals with the question about the range of
fluctuations of the random crystal around its limit shape. As we have said
above, in the 3D case the known results about the closeness of the random
crystal to its asymptotic shape are obtained only for the $L^{1}$ distance
between them, while in the 2D case they are known to hold for the Hausdorff
distance. Probably one cannot hope to extend this result to the 3D case at all
subcritical temperatures. However it is reasonable to expect that such result
does hold at very low temperatures. That was suggested already in the book
\cite{DKS}. Namely, though the solution of the Wulff variational problem is
not stable in the Hausdorff distance, due to the possibility that thin long
hairs can appear on the crystal, at low temperatures these hairs are highly
improbable due to their energetic cost. Here we give an extra reason to
believe it by proving \textquotedblleft No Hairs\textquotedblright\ theorem,
that for the low temperature SOS model in the $N\times N$ box the random
surface fluctuates away from the flat facet by less than $C\ln N,$ for some
$C<\infty,$ and so the low temperature SOS crystal is always \textquotedblleft
clean-shaven\textquotedblright.

We finish this introduction by pointing out the technical innovations of the
present paper. Usually, to prove a result of such kind, one has to obtain the
lower estimate on the probability of \textquotedblleft nice\textquotedblright%
\ behavior of the random surface we are interested in, together with the upper
estimate on the probability of its \textquotedblleft ugly\textquotedblright%
\ behavior. Here the latter is easy, while the former is very hard, since this
is the question about the typical behavior of the collection of contours which
are strongly interacting, see \cite{FPS}. We manage to establish our result by
having only the upper estimate. This is both the strong and the weak point of
our approach; we prove our theorem, but we do not have the technique to obtain
the complete control over our model.

\section{Statement of the Main Result}

Let $\varphi=\left\{  \varphi_{s}\in\mathbb{Z}\right\}  $ be an integer valued
random field, defined for $s\in\mathbb{Z}^{2}.$ Its distribution is defined by
the ``Solid-on-Solid''\ Hamiltonian
\[
H\left(  \varphi\right)  =\sum_{\substack{s,t\in\mathbb{Z}^{2}: \\\left|
s-t\right|  =1}}\left|  \varphi_{s}-\varphi_{t}\right|  .
\]
Namely, let $\Lambda\subset\mathbb{Z}^{2}$ be a finite box, $\left|
\Lambda\right|  <\infty,$ the configuration (boundary condition) $\psi_{\cdot
}$ be given outside $\Lambda,$ and the parameter $\beta>0$ (inverse
temperature) is fixed. Then we define the distribution $Q_{\beta,\Lambda,\psi
}$ on the configurations $\Omega_{\Lambda}=\left\{  \varphi_{s}:s\in
\Lambda\right\}  $ by
\begin{equation}
Q_{\beta,\Lambda,\psi}\left(  \varphi\right)  =\exp\left\{  -\beta H_{\Lambda
}\left(  \varphi\Bigm|\psi\right)  \right\}  ~/~Z\left(  \beta,\Lambda
,\psi\right)  .\label{01}%
\end{equation}
Here
\[
H_{\Lambda}\left(  \varphi\Bigm|\psi\right)  =\sum_{\substack{s\in\Lambda
,t\in\mathbb{Z}^{2}: \\\left|  s-t\right|  =1}}\left|  \left(  \varphi\vee
\psi\right)  _{s}-\left(  \varphi\vee\psi\right)  _{t}\right|  ,
\]
$\left(  \varphi\vee\psi\right)  _{t}$ equals to $\varphi_{t}$ for
$t\in\Lambda$ and to $\psi_{t}$ for $t\notin\Lambda,$ and the partition
function $Z\left(  \beta,\Lambda,\psi\right)  $ is a normalizing factor,
making $\left(  \ref{01}\right)  $ a probability distribution.

Our model of the crystal will be the distribution obtained from $Q_{\beta
,\Lambda,\psi}$ by a suitable conditioning. Namely, we will consider the case when

\begin{itemize}
\item $\Lambda=\Lambda_{N}=\left\{  s:1\leq s_{i}\leq N,i=1,2\right\}  $ is a
square $N\times N,$

\item $\psi_{\cdot}\equiv0,$

\item the volume constraint
\begin{equation}
\mathbb{V}_{N}(\varphi)=\sum_{s\in\Lambda}\varphi_{s}\geq\lambda
N^{3}\label{03}%
\end{equation}
is imposed, with $\lambda>0$ fixed.
\end{itemize}

We denote the conditional distribution $Q_{\beta,\Lambda_{N},\psi=0}\left(
\varphi\Bigm|\mathbb{V}_{N}(\varphi)\geq\lambda N^{3}\right)  $ by
$P_{\beta,N}\left(  \varphi\right)  $. We do not keep $\lambda$ in this
notation, since it will be fixed throughout the paper. We will use the
notation $Q_{\beta,N}$ for the unconditional distribution $Q_{\beta
,\Lambda_{N},\psi=0}\left(  \cdot\right)  .$ We define the crystal
$\mathsf{C}\left(  \varphi\right)  $ to be the body below the graph of
$\varphi:$%
\[
\mathsf{C}\left(  \varphi\right)  =\left\{  \left(  s,h\right)  \in
\mathbb{Z}^{3}:s\in\Lambda,0\leq h\leq\varphi\left(  s\right)  \right\}  .
\]

To formulate our results about the facets we have to introduce the level sets.
So for every $\varphi$ and every integer $l>0$ we denote by $D\left(
\varphi,l\right)  $ the subset of all sites $s$ in $\Lambda,$ where
$\varphi\left(  s\right)  \geq l.$ We identify $D\left(  \varphi,l\right)  $
with the union of the closed unit squares centered at the corresponding points
$s.$ The connected components of the topological boundary of $D\left(
\varphi,l\right)  $ will be called contours. The set of all contours will be
denoted by $\Delta\left(  \varphi,l\right)  .$ The sets $D\left(
\varphi,l\right)  $ can be disconnected; we denote by $D_{i}\left(
\varphi,l\right)  $ the collections of connected components of $D\left(
\varphi,l\right)  ,$ $i=1,2,...$ which are mutually external. They will be
called \textbf{sections}. By $\partial D_{i}\left(  \varphi,l\right)  $ we
denote the outer component of the boundary of the section $D_{i}\left(
\varphi,l\right)  $. The set of contours $\partial D_{i}\left(  \varphi
,l\right)  $ coincides with the set of external contours of the family
$\Delta\left(  \varphi,l\right)  .$ A section $D_{i}\left(  \varphi,l\right)
$ will be called \textbf{large, }if
\begin{equation}
\left|  \partial D_{i}\left(  \varphi,l\right)  \right|  \geq K\ln
N,\label{26}%
\end{equation}
where $K$ is some big constant, to be chosen later. Otherwise it is called
\textbf{small}.

Consider now the level $L=L\left(  \varphi\right)  ,$ which is defined to be
the \textbf{maximal value} of $l$-s, satisfying the following condition:

\begin{itemize}
\item $\left\vert D\left(  \varphi,l\right)  \right\vert \geq a\left(
\beta\right)  N^{2},$ where $a\left(  \beta\right)  $ is some small quantity,
$a\left(  \beta\right)  \rightarrow0$ as $\beta\rightarrow\infty,$ to be
defined later.
\end{itemize}

Denote by $\mathbf{F}_{1}\left(  \varphi\right)  $ the level set $D\left(
\varphi,L\left(  \varphi\right)  \right)  ,$ and introduce also the notation
$\mathbf{F}_{i}\left(  \varphi\right)  $ for the level sets $D\left(
\varphi,L\left(  \varphi\right)  -i+1\right)  .$ Our initial hypothesis was
that the level set $\mathbf{F}_{1}\left(  \varphi\right)  $ -- the
\textquotedblleft First Facet\textquotedblright\ -- is the facet sought, in
the sense that $\left\vert \mathbf{F}_{1}\left(  \varphi\right)  \right\vert
\geq\left(  1-a\left(  \beta\right)  \right)  N^{2}.$ However at present we
cannot prove nor disprove this statement, and we think that it is not valid.
In particular we cannot rule out the case of $\left\vert \mathbf{F}_{1}\left(
\varphi\right)  \right\vert \sim N^{2}/2,$ say. Still, we can show that a
sharply localized jump of the function $\left\vert D\left(  \varphi,l\right)
\right\vert $ happens for typical $\varphi$-s$:$

\begin{theorem}
Suppose the temperature is low enough. Then for the typical crystal
\textbf{the \textquotedblleft Second Facet\textquotedblright\ is large}:
\[
P_{\beta,N}\left\{  \varphi:\left\vert \mathbf{F}_{2}\left(  \varphi\right)
\right\vert \geq\left(  1-a\left(  \beta\right)  \right)  N^{2}\right\}
\rightarrow1\text{ as }N\rightarrow\infty,
\]
with some $a\left(  \beta\right)  \rightarrow0$ as $\beta\rightarrow\infty.$
\end{theorem}

This result means that the crystal $\mathsf{C}\left(  \varphi\right)  $ indeed
has a horizontal facet, in the following sense: the level of height $L\left(
\varphi\right)  -1$ is almost filled with sites, whereas at the levels above
$L\left(  \varphi\right)  $ only few sites belong to the crystal. We do not
know what happens at the level $L\left(  \varphi\right)  ,$ i.e. how big the
First Facet really is.

It is known that the SOS-model undergoes the roughening transition in
temperature, see \cite{FrSp}. At low temperatures the (unconstrained)
SOS-model (without condition $\left(  \ref{03}\right)  $) with zero boundary
conditions is localized, while at high temperatures it diverges
logarithmically with $N$. It is reasonable to conjecture that the roughening
temperature $T_{r}^{SOS}$ is critical for our problem as well. In particular,
it will mean that for every temperature $\beta^{-1}>T_{r}^{SOS}, $ every
$\varepsilon>0$ and for every pair $m,n$ of integers
\begin{equation}
P_{\beta,N}\left\{  \varphi:\frac{\left\vert D\left(  \varphi,L\left(
\varphi\right)  +m\right)  \right\vert }{\left\vert D\left(  \varphi,L\left(
\varphi\right)  -n\right)  \right\vert }<1-\varepsilon\right\}  \rightarrow
0\text{ as }N\rightarrow\infty,\label{66}%
\end{equation}
for any value of the parameter $a>0,$ used in the definition of the level
height $L\left(  \varphi\right)  .$ As our theorem shows, the behavior
opposite to $\left(  \ref{66}\right)  $ takes place at low temperatures, and
we conjecture that it is the case for all temperatures below $T_{r}^{SOS}.$

To formulate the No Hairs theorem we introduce the boundary $\partial
\mathbf{F}_{i}\left(  \varphi\right)  $ of the $i$-th facet to be just the
boundary $\partial D\left(  \varphi,L\left(  \varphi\right)  -i+1\right)  $.
The theorem states that inside $\partial\mathbf{F}_{2}\left(  \varphi\right)
$ the surface $\varphi$ is almost flat, up to logarithmic excitations.

\begin{theorem}
There exists $\beta_{0}$ such that for any $\beta>\beta_{0}$, one can find
$C>0$ for which the following holds
\[
P_{\beta,N}\left\{  \varphi:\ \exists s\in\mathrm{Int\,}\left(  \partial
\mathbf{F}_{2}\left(  \varphi\right)  \right)  ,\quad\left\vert \varphi
_{s}-L(\varphi)\right\vert >C\ln N\right\}  \rightarrow0\ \ \text{ as
}N\rightarrow\infty.
\]

\end{theorem}

\section{Zero temperature Ising crystal}

In this section we discuss the relations between the Conjectures 1 and 2
above. The Conjecture 2 is clearly a weaker statement, so it is not surprising
that we can prove its SOS-counterpart, while we can not prove nor disprove the
SOS-version of Conjecture 1. The real reason why the Conjecture 2 is simpler
is the fact that it is valid at zero temperature, while the Conjecture 1 is
definitely not.

The question about the shape of the crystal in the canonical Ising model
becomes in the case of zero temperature the question about the isoperimetric
problem in $\mathbb{Z}^{3}.$ Namely, we are looking into the following
problem: let $K$ be an integer, and we consider the family $\mathcal{\tilde
{V}}_{K}$ of all subsets $V\subset\mathbb{Z}^{3}$ containing precisely $K$
sites. For every $V$ we define the value $\left\vert \partial V\right\vert $
to be the number of plaquettes (of the dual lattice) in the boundary of $V;$
in other words, $\left\vert \partial V\right\vert $ is the area of the surface
$\partial V.$ We define $\mathcal{V}_{K}\subset\mathcal{\tilde{V}}_{K}$ to be
the subset consisting of \textit{minimal }$V$-s:
\[
V\in\mathcal{V}_{K}\Leftrightarrow\left\vert \partial V\right\vert =\min
_{W\in\mathcal{\tilde{V}}_{K}}\left\vert \partial W\right\vert .
\]
In the following we will not distinguish the elements of $\mathcal{V}_{K}$
which differ by translation only, thus $\mathcal{V}_{K}$ becomes a finite set,
so we can endow it with a uniform probability distribution. We want to take
$K$ to infinity and to look on the typical properties of the crystal shapes.
However what we will see depends on the values of $K.$ In case $K=M^{3}$ with
$M$ integer, the set $\mathcal{V}_{K}$ contains just one element, so the
situation is trivial. To get some interesting behavior one has to choose the
subsequence $K_{n}\rightarrow\infty$ in a special way. There are many
different options here, and we describe just one of them.

In the formulation of the theorem, which follows, the expressions
\textquotedblleft square with rounded corners\textquotedblright\ and
\textquotedblleft cube with rounded corners\textquotedblright\ are used. They
mean here the following. Let $k$ be an integer, and $Y_{1},...,Y_{4}$ be four
Young diagrams with the total number of cells less than $k.$ Then the square
$k\times k$ with four diagrams $Y_{1},...,Y_{4}$ removed from its four corners
is our \textquotedblleft square with rounded corners\textquotedblright. In the
same way, a cube $k\times k\times k$ with rounded corners is obtained from
$k$-cube by removing eight 3D Young diagrams (called also \textquotedblleft
skyscrapers\textquotedblright) $S_{1},...,S_{8},$ with the total volume below
$k,$ from its eight corners. We will call these diagrams as \textit{defects}.

Let $0<\mu<1$ be a fixed number. We take
\[
K_{n}=n^{3}+k\left(  k-1\right)  +1,
\]
where $k=\left[  \mu n\right]  $ denote the integer part of $\mu n.$

\begin{theorem}
As $n\rightarrow\infty,$ a typical random shape $V,$ drawn from the uniform
distribution on $\mathcal{V}_{K_{n}},$ can be described as follows:

$V$ is a \textquotedblleft cube with rounded corners\textquotedblright\ of
size $n,$ to one (random) face of which a monolayer is attached, which is a
\textquotedblleft square with rounded corners\textquotedblright\ of size $k.$
These roundings have asymptotic shapes as $n\rightarrow\infty$: namely, let
$x>0$ satisfies
\begin{equation}
\left(  k-4x\right)  ^{2}=\frac{2^{11}3^{3}\left(  \zeta\left(  3\right)
\right)  ^{2}}{\pi^{6}}x^{3}\label{51}%
\end{equation}
(so $x$ is of the order of $k^{2/3}$). Then each rounded corner of the square,
scaled down by a factor $x^{1/2},$ has asymptotic shape given by the Vershik
curve,
\begin{equation}
\exp\left\{  -\tfrac{\pi}{\sqrt{6}}u\right\}  +\exp\left\{  -\tfrac{\pi}%
{\sqrt{6}}v\right\}  =1,\label{50}%
\end{equation}
while each rounded corner of the cube, scaled down by a factor $\left(
k-4x\right)  ^{1/3},$ has asymptotic shape given by the Cerf-Kenyon surface,
see \cite{CK}, Theorems 1.2 and 1.3.
\end{theorem}

The curve $\left(  \ref{50}\right)  $ was obtained in \cite{VKer}. The proof
of the above result will appear later, see \cite{S2}. The equation $\left(
\ref{51}\right)  $ is related with the asymptotic numbers of partitions and
plane partitions of a large integer, see \cite{S1}, Sect. 4.1 and 4.2.

\section{\medskip Proof of the Second Facet theorem}

We first prove a weaker statement, which, in fact, contains the main part of
the proof of our result. Define $\mathbf{E}_{j}(\varphi)$ as the interior
volume of all the external contours $\partial_{i}D(\varphi,L(\varphi)-j+1) $
of the level set $D(\varphi,L(\varphi)-j+1)$.

\begin{theorem}
For any $a>0$, there is $\beta_{0}$ such that
\[
\forall\beta\geq\beta_{0},\ \ P_{\beta,N}\left\{  \varphi:\mathbf{E}%
_{2}\left(  \varphi\right)  \geq\left(  1-a\right)  N^{2}\right\}
\rightarrow1\text{ as }N\rightarrow\infty.
\]

\end{theorem}

\begin{proof}
The proof relies on energy estimates of the contours lying in the first and
second facets. For a given height configuration $\varphi$, we denote by
$\{\gamma_{i}\}_{i\leq K_{1}}$ the set of all external contours of the family
$\Delta\left(  \varphi,L(\varphi)\right)  .$ These are just the external
boundaries $\partial D_{i}(\varphi,L(\varphi))$ of the connected components of
$\mathbf{F}_{1}(\varphi)$. Similarly, $\{\Gamma_{i}\}_{i\leq K_{2}}$ will
refer to the external contours in $\mathbf{F}_{2}(\varphi)$. By construction
the contours satisfy a compatibility condition, namely for any $\gamma_{i}$
there exists $\Gamma_{j}$ such that $\gamma_{i}$ lies inside $\Gamma_{j}$.

We introduce two events; the first one, $\mathcal{S},$ consists of
configurations such that the volume contribution to $\mathbf{E}_{1}(\varphi) $
of external small contours of the first facet is larger than $\frac{a}{2}%
N^{2},$ while the second, $\mathcal{L},$ corresponds to the configurations for
which the volume of the external large contours in the first facet is above
$\frac{a}{2}N^{2}$, and also the volume of the external large contours in the
second facet is smaller than $\left(  1-a\right)  N^{2}:$
\begin{align}
\mathcal{S}  & =\left\{  \varphi;\ \ \sum_{\gamma_{i}\ \mathrm{small}%
}|\mathrm{Int\,}(\gamma_{i})|\geq\frac{a}{2}N^{2}\right\}  \,,\label{21}\\
\mathcal{L}  & =\left\{  \varphi;\ \ \sum_{\gamma_{i}\ \mathrm{large}%
}|\mathrm{Int}(\gamma_{i})|\geq\frac{a}{2}N^{2},\ \ \sum_{\Gamma
_{j}\ \mathrm{large}}|\mathrm{Int}(\Gamma_{j})|<(1-a)N^{2}\right\}
\,.\label{34}%
\end{align}

By construction, $\left\vert \mathbf{E}_{1}\left(  \varphi\right)  \right\vert
\geq\left\vert \mathbf{F}_{1}\left(  \varphi\right)  \right\vert \geq aN^{2},$
so we can write
\[
P_{\beta,N}\left\{  \varphi:\left\vert \mathbf{E}_{2}\left(  \varphi\right)
\right\vert <\left(  1-a\right)  N^{2}\right\}  \leq P_{\beta,N}%
(\mathcal{S})+P_{\beta,N}(\mathcal{L})\,.
\]
Thus, to complete the proof, it is enough to show that for $\beta$ large
enough, there exists $c=c(a,\beta)>0$ such that
\begin{equation}
P_{\beta,N}(\mathcal{L})\leq\exp(-cN)\,,\label{22}%
\end{equation}
\begin{equation}
P_{\beta,N}(\mathcal{S})\leq\exp\left(  -c\left(  \frac{N}{\ln N}\right)
^{2}\right)  \,.\label{23}%
\end{equation}

The estimate $\left(  \ref{22}\right)  $ on the large contours will be
obtained in Subsection \ref{subsec: large contours} and the estimate $\left(
\ref{23}\right)  $ on the small contours in Subsection
\ref{subsec: small contours}.
\end{proof}

\subsection{A priory estimates on the height of the facet}

\label{subsec: height of the facet}

We start with very elementary estimates. Every SOS-surface $\varphi\in
\Omega_{\Lambda_{N}}$ is made from $N^{2}$ horizontal plaquettes and a number
of vertical ones; we denote this last one by $S\left(  \varphi\right)  $.
Evidently, the distribution $Q_{\beta,N}(\varphi)$ equals to $\exp\left\{
-\beta S\left(  \varphi\right)  \right\}  ,$ up to normalization factor.
Standard Peierls and counting arguments lead to the following simple
estimate:
\begin{equation}
Q_{\beta,N}(\varphi:S\left(  \varphi\right)  =S)\leq3^{N^{2}+S}\exp\left\{
-\beta S\right\}  .\label{24}%
\end{equation}

Under $P_{\beta,N}$, the facet should be located with a high probability at a
height of order $N$. But for us a weaker statement will be sufficient. For $K$
$>1$ and $k>0$ we define
\[
\mathcal{H}=\left\{  \varphi;~k\leq L(\varphi)\leq KN\right\}  \,.
\]

\begin{proposition}
\label{prop: height} For any $\beta$ large enough, any $k$ fixed and $K\geq
K\left(  \beta\right)  $ large enough
\begin{equation}
P_{\beta,N}\left(  \mathcal{H}\right)  \geq1-\exp(-\beta^{\prime}%
N^{2})\,,\label{25}%
\end{equation}
where $\beta^{\prime}$ diverges with $\beta$.
\end{proposition}

\begin{proof}
Our claim follows easily from $\left(  \ref{24}\right)  .$ Indeed, the
property $L(\varphi)>KN$ implies that
\[
S\left(  \varphi\right)  \geq4K\sqrt{a\left(  \beta\right)  }N^{2},
\]
since for every level $l$ below $L(\varphi)$ we have $\left\vert D\left(
\varphi,l\right)  \right\vert \geq a\left(  \beta\right)  N^{2}.$ On the other
hand, if the surface $\varphi$ is such that all its sections $D_{i}\left(
\varphi,l\right)  $ have the area below $a\left(  \beta\right)  N^{2}$ for
$l\geq k,$ then
\[
S\left(  \varphi\right)  \geq\frac{\lambda N^{3}-kN^{2}}{a\left(
\beta\right)  N^{2}}4\sqrt{a\left(  \beta\right)  }N=C\left(  \beta\right)
N^{2},\ \text{with }C\left(  \beta\right)  \rightarrow\infty\text{ as }%
\beta\rightarrow\infty.
\]
Therefore
\[
P_{\beta,N}\left\{  \varphi:L(\varphi)\notin\left[  k,KN\right]  \right\}
\leq\frac{Q_{\beta,N}(S(\varphi)\geq\tilde{C}\left(  \beta\right)  N^{2}%
)}{Q_{\beta,N}({\ }\mathbb{V}_{N}\geq\lambda N^{3})}\leq\exp\left(
-\beta^{\prime}N^{2}\right)  \,,
\]
since the denominator is always larger than $\exp(-\beta\lambda N^{2}).$
\end{proof}

\subsection{Isoperimetric inequality - zero temperature case}

Here we prove the statement mentioned in the introduction, that merging two
droplets together decreases the surface energy. More generally, just
increasing the bigger one at the expense of the smaller one still makes the
energy smaller. We prove here the corresponding statement for the 2D zero
temperature Ising model only, while some generalizations are available by
using the technique of \cite{SS1}.

For an integer $V$ we define $L=L\left(  V\right)  $ to be the largest integer
such that $L^{2}\left(  V\right)  \leq V,$ and we introduce $r=r\left(
V\right)  =V-L^{2}\left(  V\right)  .$ We denote by $p=p\left(  V\right)  $
the length of the shortest closed path on the lattice $\mathbb{Z}^{2},$
surrounding $V$ unit plaquettes. Clearly,
\[
p\left(  V\right)  =\left\{
\begin{array}
[c]{ll}%
4L\left(  V\right)   & \text{ if }V=L^{2}\left(  V\right)  ,\\
4L\left(  V\right)  +2 & \text{ if }0<r\left(  V\right)  \leq L\left(
V\right)  ,\\
4L\left(  V\right)  +4 & \text{ if }L\left(  V\right)  <r\left(  V\right)
\leq2L\left(  V\right)  .
\end{array}
\right.
\]
We will call $p\left(  V\right)  $ the surface energy of the droplet $V.$ In
what follows we will identify the integers $V$ with the collections of
plaquettes from $\mathbb{Z}^{2}$ with perimeter $p\left(  V\right)  ,$ which
will be called also droplets. Now, let $V_{1}\leq V_{2}$ be two integers, and
we suppose that for some $N$ and some (small) $\rho>0$ we have
\[
V_{1}\geq\rho N^{2},
\]%
\[
V_{2}\leq N^{2}.
\]
The second condition means that the larger droplet $V_{2}$ fits inside the
volume $N\times N,$ and the first one -- that the smaller droplet is not too small.

Let now $D$ be any integer, satisfying the conditions
\[
V_{1}\geq D\geq\rho N^{2}.
\]

\begin{lemma}
\label{drop} The transfer of the amount $D$ from the droplet $V_{1}$ to
$V_{2}$ decreases the total surface energy: there exists a constant
$\varkappa=\varkappa\left(  \rho\right)  >0,$ such that
\[
\left(  1-\varkappa\left(  \rho\right)  \right)  \left(  p\left(
V_{1}\right)  +p\left(  V_{2}\right)  \right)  \geq p\left(  V_{1}-D\right)
+p\left(  V_{2}+D\right)  .
\]

\end{lemma}

\begin{proof}
We will show that the difference
\[
p\left(  V_{1}-D\right)  +p\left(  V_{2}+D\right)  -p\left(  V_{1}\right)
-p\left(  V_{2}\right)
\]
is of the order of $p\left(  V_{1}\right)  +p\left(  V_{2}\right)  $ and
negative. Since the function $p\left(  x\right)  $ equals approximately to
$4\sqrt{x}$ -- more precisely,
\begin{equation}
4\sqrt{x}\leq p\left(  x\right)  <4\sqrt{x}+4,\label{40}%
\end{equation}
-- it is enough to show that the difference
\[
\sqrt{V_{1}-D}+\sqrt{V_{2}+D}-\sqrt{V_{1}}-\sqrt{V_{2}}%
\]
is of the order of $\sqrt{V_{1}}+\sqrt{V_{2}}$ and negative. Let us rewrite
the difference as
\[
\sqrt{V_{1}}\left(  \sqrt{1-\frac{D}{V_{1}}}-1\right)  +\sqrt{V_{2}}\left(
\sqrt{1+\frac{D}{V_{2}}}-1\right)
\]
and use the Taylor expansion of the function $\sqrt{1+x}.$ We get
\begin{align}
& \sqrt{V_{1}}\left(  \sqrt{1-\frac{D}{V_{1}}}-1\right)  +\sqrt{V_{2}}\left(
\sqrt{1+\frac{D}{V_{2}}}-1\right)  \nonumber\\
& =\frac{1}{2}\left(  -\frac{D}{\sqrt{V_{1}}}+\frac{D}{\sqrt{V_{2}}}\right)
-\frac{1}{8}\left(  \frac{D^{2}}{V_{1}\sqrt{V_{1}}}+\frac{D^{2}}{V_{2}%
\sqrt{V_{2}}}\right)  +...\label{48}%
\end{align}
Now, since $V_{1}<V_{2},$ the contents of all the odd brackets are negative,
while the even coefficients are also negative, so the difference is negative
as well. Finally, since all the values $V_{1},V_{2}$ and $D$ are of the same
order, the second term $\frac{D^{2}}{V_{1}\sqrt{V_{1}}}+\frac{D^{2}}%
{V_{2}\sqrt{V_{2}}}$ is of the order of $\sqrt{V_{1}}+\sqrt{V_{2}},$ and the
proof follows.
\end{proof}

\subsection{Estimates on the large contours}

\label{subsec: large contours}

In this section we will prove the estimate $\left(  \ref{22}\right)  .$

For a given integer $\ell\geq2$ and a compatible collection of large contours
$(\gamma,\Gamma)=(\left\{  \gamma_{i}\right\}  ,\left\{  \Gamma_{j}\right\}
)$ we denote by $\varphi\sim(\gamma,\Gamma,\ell)$ the height configurations
$\varphi$ which satisfy:

\begin{itemize}
\item the volume constraint ${\ }\mathbb{V}_{N}(\varphi)=\sum_{s\in\Lambda
}\varphi_{s}\geq\lambda N^{3},$

\item $L(\varphi)=\ell,$

\item the only exterior large contours on the level sets $\mathbf{F}_{1}$ and
$\mathbf{F}_{2}$ are given by $(\gamma,\Gamma)$.
\end{itemize}

Then we have%

\[
P_{\beta,N}(\mathcal{L})\leq P_{\beta,N}(\mathcal{H}^{c})+\frac{1}{Q_{\beta
,N}({\ }\mathbb{V}_{N}\geq\lambda N^{3})}\sum_{\ell=2}^{KN}\ \ \sum
_{(\gamma,\Gamma)\in\mathcal{L}}\ \ \sum_{\varphi\sim(\gamma,\Gamma,\ell
)}\ Q_{\beta,N}(\varphi)\,.
\]

For a given triplet $(\gamma,\Gamma,\ell)$, we define the erasing map
\begin{equation}
\varphi\rightsquigarrow\hat{\varphi}=\left(  \hat{\varphi}_{s}=\varphi
_{s}-\sum_{i}1_{\{s\in\mathrm{Int}(\gamma_{i})\}}-\sum_{j}1_{\{s\in
\mathrm{Int}(\Gamma_{j})\}}\right)  _{s\in\Lambda}\,.\label{44}%
\end{equation}
It maps injectively the set $\{\varphi\sim(\gamma,\Gamma,\ell)\}$ into the
set
\[
\left\{  {\ }\mathbb{V}_{N}(\varphi)\geq\lambda N^{3}-\sum_{i}|\mathrm{Int}%
(\gamma_{i})|-\sum_{j}|\mathrm{Int}(\Gamma_{j})|\right\}  \,.
\]
Evidently,
\[
Q_{\beta,N}(\varphi)=\exp\big(-\beta\sum_{i}|\gamma_{i}|-\beta\sum_{j}%
|\Gamma_{j}|\big)\ Q_{\beta,N}(\hat{\varphi})\,.
\]
Therefore we have the \textquotedblleft Peierls estimate\textquotedblright\
\begin{align}
P_{\beta,N}(\mathcal{L}) &  \leq P_{\beta,N}(\mathcal{H}^{c})+\sum_{\ell
=2}^{KN}\ \ \sum_{(\gamma,\Gamma)\in\mathcal{L}}\exp\big(-\beta\sum_{i}%
|\gamma_{i}|-\beta\sum_{j}|\Gamma_{j}|\big)\label{30}\\
&  \qquad\qquad\qquad\times\frac{Q_{\beta,N}\big({\ }\mathbb{V}_{N}%
(\varphi)\geq\lambda N^{3}-\sum_{i}|\mathrm{Int}(\gamma_{i})|-\sum
_{j}|\mathrm{Int}(\Gamma_{j})|\big)}{Q_{\beta,N}\big({\ }\mathbb{V}%
_{N}(\varphi)\geq\lambda N^{3}\big)}\,.\nonumber
\end{align}

The important quantity is the total volume of the interiors of the contours,
thus we are going to average on all the possible contour shapes in order to
retain only the information on the volume. Fix $V\in\lbrack\frac{a}{2}%
N^{2},(1-a)N^{2}]$ and consider the collection $(\gamma_{i})$ of the large
contours such that
\begin{equation}
\sum_{i}|\mathrm{Int}(\gamma_{i})|=V.\label{27}%
\end{equation}
The optimal shape for a contour of volume $V$ is a square of side length
$L=L\left(  V\right)  $ with (possibly) an additional layer of $r\left(
V\right)  $ sites such that
\[
V=L\left(  V\right)  ^{2}+r\left(  V\right)  ,\qquad r\left(  V\right)
\in\{0,\dots,2L\left(  V\right)  \}\,.
\]
In this case the following isoperimetric inequality holds uniformly over the
collection $(\gamma_{i})$ which satisfy the volume constraint $\left(
\ref{27}\right)  :$
\[
\sum_{i}|\gamma_{i}|\geq4L\left(  V\right)  \,.
\]

Let $\beta^{\prime}=\beta-10$. Summing over all the collections of contours
such that $\sum_{i}|\mathrm{Int}(\gamma_{i})|=V$, we get
\begin{align*}
\sum_{(\gamma_{i})}\exp\big(-\beta\sum_{i}|\gamma_{i}|\big)  & \leq\exp
(-\beta^{\prime}4L\left(  V\right)  )\left(  \sum_{(\gamma_{i})}\prod_{i}%
\exp\big(-10|\gamma_{i}|\big)\right) \\
& \leq\exp(-4\beta^{\prime}L\left(  V\right)  )\,,
\end{align*}
where the final inequality is obtained by taking into account the entropy of a
single large contour
\[
\sum_{(\gamma_{i})}\prod_{i}\exp\big(-10|\gamma_{i}|\big)\leq\left(
1+\sum_{\ell\geq K\ln N}3^{\ell}\exp(-10\ell)\right)  ^{N^{2}}-1<1\,.
\]

Plugging this inequality in $\left(  \ref{30}\right)  $, we get
\begin{align}
P_{\beta,N}(\mathcal{L}) &  \leq P_{\beta,N}(\mathcal{H}^{c})\nonumber\\
&  +\sum_{\ell=2}^{N^{3}}\ \ \sum_{\sqrt{\frac{a}{2}}N\leq L_{1}\leq L_{2}%
\leq\sqrt{1-a}N}(2L_{1}+1)(2L_{2}+1)\exp\big(-4\beta^{\prime}(L_{1}%
+L_{2})\big)\label{31}\\
&  \qquad\qquad\times\frac{Q_{\beta,N}\big({\ }\mathbb{V}_{N}(\varphi
)\geq\lambda N^{3}-(L_{1}+1)^{2}-(L_{2}+1)^{2}\big)}{Q_{\beta,N}%
\big(\mathbb{V}_{N}(\varphi)\geq\lambda N^{3}\big)}\,.\nonumber
\end{align}
Here we have indexed the volume $V_{1}$ of the large contours in
$\mathbf{F}_{1}(\varphi)$ by the parameter $L_{1}$ such that $V_{1}=L_{1}%
^{2}+r_{1}$ (with $0\leq r_{1}\leq2L_{1}$). Thus for a given $L_{1}$, there is
at most $(2L_{1}+1)$ corresponding quasi-cubes. Similarly, the contours in
$\mathbf{F}_{2}$ are indexed by $L_{2}$.

The final step is to show that for $\sqrt{\frac{a}{2}}N\leq L_{1}\leq
L_{2}\leq\sqrt{1-a}N$ we have
\begin{equation}
\frac{Q_{\beta,N}\big({\ }\mathbb{V}_{N}(\varphi)\geq\lambda N^{3}%
-(L_{1}+1)^{2}-(L_{2}+1)^{2}\big)}{Q_{\beta,N}\big({\ }\mathbb{V}_{N}%
(\varphi)\geq\lambda N^{3}\big)}\leq\exp\big(4(\beta^{\prime}-10)(L_{1}%
+L_{2})\big)\,.\label{32}%
\end{equation}
This inequality combined with $\left(  \ref{31}\right)  $ will imply $\left(
\ref{22}\right)  $.

For this we will use our Lemma \ref{drop}, with $D=\min\left\{  (L_{1}%
+1)^{2},N^{2}-(L_{2}+1)^{2}\right\}  .$ From it we know that for $\beta$ and
$N$ large enough
\begin{equation}
4(1-\frac{20}{\beta})\big(L_{1}+L_{2}\big)\geq p\left(  (L_{1}+1)^{2}%
-D\right)  +p\left(  (L_{2}+1)^{2}+D\right)  .\label{33}%
\end{equation}
As a consequence the function $a\left(  \beta\right)  $ of the Theorem 3 must
be chosen in such a way that
\begin{equation}
\varkappa\left(  a\left(  \beta\right)  \right)  >20/\beta.\label{67}%
\end{equation}
Any choice of the function $a\left(  \beta\right)  \rightarrow0$ as
$\beta\rightarrow\infty,$ consistent with $\left(  \ref{67}\right)  ,$ is
allowed in Theorem 3. So $a\left(  \beta\right)  $ vanishes as $\beta$
diverges, but nevertheless it cannot be too small. The heuristic reason is
that for any finite $\beta$ the macroscopic crystal has rounded edges; thus
there exists a constant $a^{\prime}\left(  \beta\right)  >0$ such that it is
no longer favorable to erase level sets which have the volume larger than
$\left(  1-a^{\prime}\left(  \beta\right)  \right)  N^{2}$.

Now to any height configuration $\varphi$ in $\{\mathbb{V}_{N}(\varphi
)\geq\lambda N^{3}-(L_{1}+1)^{2}-(L_{2}+1)^{2}\}$, we associate the
configuration $\check{\varphi},$ defined by
\begin{equation}
\check{\varphi}_{s}=\varphi_{s}+1_{\{s\in\left[  (L_{2}+1)^{2}+D\right]
\}}+1_{\{s\in\left[  (L_{1}+1)^{2}-D\right]  \}}\,,~\forall s=(s_{1},s_{2}%
)\in\Lambda,\label{63}%
\end{equation}
where $\left[  n\right]  $ denotes the square droplet $n\times n.$ Here one
needs of course to fix the position of the smaller droplet $\left[
(L_{1}+1)^{2}-D\right]  $ to be inside the larger one, $\left[  (L_{2}%
+1)^{2}+D\right]  .$ This however holds automatically, since either the
droplet $\left[  (L_{1}+1)^{2}-D\right]  $ is empty, or $\left[  (L_{2}%
+1)^{2}+D\right]  =N\times N.$

The correspondence $\varphi\rightsquigarrow\check{\varphi}$ maps injectively
the set $\{\mathbb{V}_{N}(\varphi)\geq\lambda N^{3}-(L_{1}+1)^{2}%
-(L_{2}+1)^{2}\}$ into $\{\mathbb{V}_{N}(\varphi)\geq\lambda N^{3}\}$.
Furthermore, the energy difference between the height configurations $\varphi$
and $\check{\varphi}$ is bounded by $p\left(  (L_{1}+1)^{2}-D\right)
+p\left(  (L_{2}+1)^{2}+D\right)  $, so
\begin{align*}
& \frac{Q_{\beta,N}\big(\varphi:\mathbb{V}_{N}(\varphi)\geq\lambda
N^{3}-(L_{1}+1)^{2}-(L_{2}+1)^{2}\big)}{Q_{\beta,N}\big({\ }\varphi
:\mathbb{V}_{N}(\varphi)\geq\lambda N^{3}\big)}\\
& \leq\exp\left\{  \beta\left[  p\left(  (L_{1}+1)^{2}-D\right)  +p\left(
(L_{2}+1)^{2}+D\right)  \right]  \right\}  \,.
\end{align*}
Combining this inequality with $\left(  \ref{33}\right)  $, we conclude that
$\left(  \ref{32}\right)  $ holds.

\vskip.4cm

\subsection{Estimates on the small contours}

\label{subsec: small contours}

This subsection contains the proof of the estimate $\left(  \ref{23}\right)
.$

We follow the scheme of the proof used to control the phase of small contours
in the Ising model (see \cite{SS2}). Define the subset $\Lambda^{(0)}=\left(
K_{N}\times\mathbb{Z}\right)  ^{2}\cap\Lambda$, where $K_{N}=2K\ln N$. For any
site $s$ in $\Lambda$ such that $\Vert s\Vert_{\infty}<K_{N}$, the shift of
$\Lambda^{(0)}$ by $s$ is denoted by $\Lambda^{(s)}$. To any collection of
contours $\gamma=(\gamma_{i})$, the number of sites in $\Lambda^{(s)}$
belonging to the interior of a contour in $\gamma$ is denoted by
$\mathcal{N}_{s}$.

If $\varphi$ belongs to $S$ and $\gamma$ is the collection of all the small
contours in $\mathbf{F}_{1}(\varphi)$, then $\sum_{\Vert s\Vert_{\infty}\leq
K_{N}}\mathcal{N}_{s}\geq\frac{a}{2}N^{2}\,.$ Thus there exists at least one
site $s$ such that
\begin{equation}
\mathcal{N}_{s}\geq\frac{a}{2}\left(  \frac{N}{K_{N}}\right)  ^{2}%
\,.\label{38}%
\end{equation}
Denote by $\Delta^{s}$ the set of collections of exterior small contours
$\gamma^{(s)}=(\gamma_{i}^{(s)})$, such that the condition $\left(
\ref{38}\right)  $ is fulfilled. Now for $\gamma^{(s)}\in\Delta^{s}$ and
$\ell\geq2$ we introduce the sets $\mathcal{S}\left(  \gamma^{(s)}%
,\ell\right)  $ of the height configurations $\varphi,$ which satisfy the properties:

\begin{itemize}
\item $L(\varphi)=\ell,~\mathbb{V}_{N}(\varphi)\geq\lambda N^{3},$

\item the contours from $\gamma^{(s)}$ are among the exterior contours of
$\mathbf{F}_{1}(\varphi).$
\end{itemize}

We get
\[
P_{\beta,N}(\mathcal{S})\leq P_{\beta,N}(\mathcal{H}^{c})+\sum_{\ell=2}%
^{N^{3}}\ \ \sum_{\Vert s\Vert_{\infty}\leq K_{N}}\ \ \sum_{\gamma^{(s)}%
\in\Delta^{s}}\ \ \sum_{\varphi\sim\mathcal{S}\left(  \gamma^{(s)}%
,\ell\right)  }\ P_{\beta,N}(\varphi)\,.
\]

We proceed as in the previous subsection and erase all the small contours.
belonging to the set $\gamma^{(s)}$. The total volume contribution of these
contours is always smaller than $N^{2}$, so we get
\begin{align}
P_{\beta,N}(\mathcal{S}) &  \leq P_{\beta,N}(\mathcal{H}^{c})+\sum_{\ell
=2}^{N^{3}}\ \ \sum_{\Vert s\Vert_{\infty}\leq K_{N}}\ \ \sum_{\gamma^{(s)}%
\in\Delta^{s}}\ \ \exp\big(-\beta\sum_{i}|\gamma_{i}^{(s)}|\big)\label{47}\\
&  \qquad\qquad\times\frac{Q_{\beta,N}\big(\varphi:{\ }\mathbb{V}_{N}%
(\varphi)\geq\lambda N^{3}-N^{2}\big)}{Q_{\beta,N}\big(\varphi:{\ }%
\mathbb{V}_{N}(\varphi)\geq\lambda N^{3}\big)}.\nonumber
\end{align}

Using a shift of the height configurations by 1, we easily see that
\[
\frac{Q_{\beta,N}\big(\mathbb{V}_{N}(\varphi)\geq\lambda N^{3}-N^{2}%
\big)}{Q_{\beta,N}\big(\mathbb{V}_{N}(\varphi)\geq\lambda N^{3}\big)}\leq
\exp(4\beta N)\,.
\]

In order to complete the derivation of $\left(  \ref{23}\right)  $, it is
enough to prove that there is $c>0$ such that uniformly in $s$,
\begin{equation}
\sum_{\gamma^{(s)}\in\Delta^{s}}\ \exp\big(-\beta\sum_{i}|\gamma_{i}%
^{(s)}|\big)\leq\exp\left(  -cN^{2}/K_{N}^{2}\right)  \,.\label{41}%
\end{equation}

The occurrence of a small contour surrounding a site $i_{0}$ is bounded by
\[
\sum_{\Gamma\ni i_{0}}\exp(-\beta|\Gamma|)\leq\sum_{\ell\geq4}\ell3^{\ell}%
\exp(-\beta\ell)=q_{\beta}\,,
\]
where $q_{\beta}$ vanishes as $\beta$ goes to infinity.

Denote the number of sites $N^{2}/K_{N}^{2}$ in $\Lambda^{(s)}$ by $M_{N}$.
The small contours surrounding different sites of $\Lambda^{(s)}$ are
independent, so we will obtain the bound $\left(  \ref{41}\right)  $ from an
upper bound for a large deviation of a system of $M_{N}$ independent random
variables. We have
\begin{align*}
\sum_{\gamma^{(s)}\in\Delta^{s}}\exp\left(  -\beta\sum_{i}|\gamma_{i}%
^{(s)}|\right)   & \leq\sum_{k\geq\frac{a}{2}M_{N}}\binom{M_{N}}{k}q_{\beta
}^{k}\\
& \leq\left(  \frac{1}{1-q_{\beta}}\right)  ^{M_{N}}\sum_{k\geq\frac{a}%
{2}M_{N}}\binom{M_{N}}{k}q_{\beta}^{k}\left(  1-q_{\beta}\right)  ^{M_{N}-k},
\end{align*}
where the last sum is the probability of the following event:

Let $\xi_{1},...,\xi_{M_{N}}$ be i.i.d. random variables, taking values $1$
with probability $q_{\beta}$ and $0$ with probability $1-q_{\beta}.$ Then
$\sum_{k\geq\frac{a}{2}M_{N}}\binom{M_{N}}{k}q_{\beta}^{k}\left(  1-q_{\beta
}\right)  ^{M_{N}-k}=\Pr\left\{  \xi_{1}+...+\xi_{M_{N}}\geq\frac{a}{2}%
M_{N}\right\}  .$ It is well known that such probability can be estimated from
above by $\exp(-c_{\beta}M_{N}),$ where $c_{\beta}$ is a positive constant for
$\beta$ large enough. For a reference one can consult, for example, Lemma 10
and Corollary 11 of \cite{MRSV}. On the other hand $q_{\beta}\rightarrow0$ as
$\beta\rightarrow\infty,$ so the estimate $\left(  \ref{41}\right)  $ follows.

\subsection{End of the proof}

\label{end}

Thus far we have proven that the quantity $\mathbf{E}_{2}(\varphi)$ -- the
area of the external contours $\{\Gamma_{i}\}_{i\leq K_{2}}$ of the second
facet $\mathbf{F}_{2}(\varphi)$ -- is typically above the level $\left(
1-a\left(  \beta\right)  \right)  N^{2}.$ We are going to explain now that
this in fact easily implies that the area of the facet $\mathbf{F}_{2}%
(\varphi)$ itself has to be above the level $\left(  1-2a\left(  \beta\right)
\right)  N^{2}.$ Indeed, suppose that two events happen:
\begin{equation}
\mathbf{E}_{2}\left(  \varphi\right)  \geq\left(  1-a\left(  \beta\right)
\right)  N^{2},\text{ and }\left\vert \mathbf{F}_{2}\left(  \varphi\right)
\right\vert \leq\left(  1-2a\left(  \beta\right)  \right)  N^{2}.\label{42}%
\end{equation}
We will show that its probability vanishes as $N\rightarrow\infty.$

To see this, let us introduce the \textit{second order }external contours
$\left\{  \tilde{\Gamma}_{j}\right\}  _{j\leq\tilde{K}_{2}},$ by defining them
to be all the external contours of the collection

\noindent$\Delta\left(  \varphi,L(\varphi)-1\right)  ~\backslash~\{\Gamma
_{i}\}_{i\leq K_{2}}.$ (So the contours $\{\Gamma_{i}\}_{i\leq K_{2}}$ should
be called the \textit{first order }external contours.)

Under $\left(  \ref{42}\right)  $ we have that
\[
\sum_{j\leq\tilde{K}_{2}}|\mathrm{Int}(\tilde{\Gamma}_{j})|\geq aN^{2}.
\]
Note also that the erasing map is now given by
\[
\varphi\rightsquigarrow\check{\varphi}=\left(  \check{\varphi}_{s}=\varphi
_{s}+\sum_{j\leq\tilde{K}_{2}}1_{\{s\in\mathrm{Int}(\tilde{\Gamma}_{j}%
)\}}\right)  _{s\in\Lambda}\,,
\]
(compare with $\left(  \ref{44}\right)  $), so
\begin{equation}
\mathbb{V}_{N}(\check{\varphi})\geq\mathbb{V}_{N}(\varphi)\geq\lambda
N^{3}.\label{45}%
\end{equation}

As above, we split the second order\textit{\ }external contours $\left\{
\tilde{\Gamma}_{j}\right\}  _{j\leq\tilde{K}_{2}}$ into small and large ones.
We introduce two events; the first one, $\mathcal{\tilde{S}},$ consists of
configurations such that their small second order external contours of the
second facet have a total volume larger than $\frac{a}{2}N^{2},$ while the
second, $\mathcal{\tilde{L}},$ corresponds to the configurations with the
volume of the large second order external contours of the second facet is
above $\frac{a}{2}N^{2}:$
\begin{align*}
\mathcal{\tilde{S}}  & =\left\{  \varphi;\ \ \sum_{\tilde{\Gamma}%
_{j}\ \mathrm{small}}|\mathrm{Int}(\tilde{\Gamma}_{j})|\geq\frac{a}{2}%
N^{2}\right\}  \,,\\
\mathcal{\tilde{L}}  & =\left\{  \varphi;\ \ \sum_{\tilde{\Gamma}%
_{j}\ \mathrm{large}}|\mathrm{Int}(\tilde{\Gamma}_{j})|\geq\frac{a}{2}%
N^{2}\right\}  \,.
\end{align*}
Then we can estimate the probabilities $P_{\beta,N}(\mathcal{\tilde{L}%
}),\ P_{\beta,N}(\mathcal{\tilde{S}})$ by repeating the estimates for
$P_{\beta,N}(\mathcal{L}),\ P_{\beta,N}(\mathcal{S}),$ obtained above. In
fact, the corresponding estimates are even simpler, because the analogs of the
estimates $\left(  \ref{31}\right)  ,$ $\left(  \ref{47}\right)  ,$ do not
contain the factors $\frac{Q_{\beta,N}\big(\varphi:{\ }\mathbb{V}_{N}%
(\varphi)\geq\lambda N^{3}-\cdot\big)}{Q_{\beta,N}\big(\varphi:{\ }%
\mathbb{V}_{N}(\varphi)\geq\lambda N^{3}\big)},$ due to $\left(
\ref{45}\right)  .$

\section{Proof of the No Hairs theorem}

The excitations of the microscopic crystal around the second facet will be
called hairs. There are two kinds of hairs: the up-hairs and the down-hairs.
The up-hair of the SOS-surface $\varphi$ is a sequence $\gamma_{1}%
,\dots,\gamma_{H}$ of contours, such that:

\begin{itemize}
\item
\begin{equation}
\gamma_{i}\in\partial D\left(  \varphi,L\left(  \varphi\right)  +i\right)
,\label{60}%
\end{equation}

\item contours $\gamma_{i}$ are ordered by inclusion, which means that
$\mathrm{Int\,}\left(  \gamma_{i+1}\right)  \subseteq\mathrm{Int\,}\left(
\gamma_{i}\right)  $ for all $i\geq1,$

\item $\mathrm{Int}\left(  \gamma_{1}\right)  \subseteq\mathrm{Int\,}\left(
\partial\mathbf{F}_{2}\left(  \varphi\right)  \right)  ,$

\item the sequence $\gamma_{1},\dots,\gamma_{H}$ is maximal, in the sense that
there is no longer sequence of external contours, satisfying all of the above,
of which our sequence is a subsequence.
\end{itemize}

We denote such an up-hair by $\Gamma=\{\gamma_{1},\dots,\gamma_{H}\}.$

The down-hair is defined as a similar sequence $\hat{\Gamma}=\{\hat{\gamma
}_{1},\dots,\hat{\gamma}_{H}\}$ of contours, except that in $\left(
\ref{60}\right)  $ the sign is opposite: $\hat{\gamma}_{i}\in\partial D\left(
\varphi,L\left(  \varphi\right)  -i\right)  .$ The value $H$ will be called
the length of the hair.

Clearly, our statement is equivalent to proving that the probability of
occurrence of a hair with length $H>C\ln N$ vanishes in the limit as
$N\rightarrow\infty.$ We denote such an event by $A_{C}.$

In what follows we will treat only the up-hairs, since the case of the
down-hairs is simpler, as there is no volume constraint (see the argument of
the subsection \ref{end}); therefore, in the rest of this section we will use
the term \textquotedblleft hair\textquotedblright\ instead of
\textquotedblleft up-hair\textquotedblright.

By our definitions we have that $|$Int$(\gamma_{1})|\leq aN^{2}.$ We introduce
now the sequence
\[
v_{r}(N)=a\frac{N^{2}}{2^{r}},~r=1,2,...,
\]
and we will characterize each hair by the amounts $H_{r}(\Gamma)$ of its
contours $\gamma_{i}$ such that
\begin{equation}
v_{r+1}(N)<|\mathrm{Int\,}\left(  \gamma_{i}\right)  |\leq v_{r}%
(N)\,.\label{61}%
\end{equation}
Naturally, we need only these $v_{r}(N)$-s which are $\geq1$, so we define
$R_{N}$ to be the largest value of $r$ such that the scale $v_{r}(N)\geq1$. We
fix also the intermediate scale $R_{N}^{\prime}<R_{N}:$%
\[
R_{N}^{\prime}=\frac{1}{\ln2}\ln\left(  \frac{aN^{2}}{C_{1}(\ln N)^{2}%
}\right)  .
\]
Then for every $r\leq R_{N}^{\prime}$ we have
\[
v_{r}(N)\geq C_{1}(\ln N)^{2}\,,
\]
and the choice of the constant $C_{1}$ is made in such a way that the contours
of volume larger than $C_{1}(\ln N)^{2}$ are large contours, i.e. their
perimeter is larger than $K\ln N$.

Finally we introduce the sequence $h_{r}(N),$ defined as follows:
\[
h_{r}(N)=\left\{
\begin{array}
[c]{ll}%
4, & \text{ if ~}r<R_{N}^{\prime}\,,\\
C_{2}\,2^{r/2}N^{-1}\ln N, & \text{ if ~}r\geq R_{N}^{\prime}\,,
\end{array}
\right.
\]
where $C_{2}$ is chosen such that
\begin{equation}
h_{R_{N}^{\prime}}(N)=\frac{\sqrt{a}\,C_{2}\ln N}{\sqrt{v_{R_{N}^{\prime}}%
(N)}}=\frac{\sqrt{a}}{\sqrt{C_{1}}}C_{2}\geq10\,.\label{64}%
\end{equation}
By definition,
\[
\sum_{r=0}^{R_{N}}h_{r}(N)\leq4R_{N}^{\prime}+\sum_{r=R_{N}^{\prime}}^{R_{N}%
}C_{2}2^{r/2}N^{-1}\ln N\leq C_{3}\ln N\,,
\]
for some $C_{3}>0$.

Let us fix the constant $C$ of our theorem to be much larger than $C_{3}$. If
the length of $\Gamma$ exceeds $C\ln N$, then we define the value $r_{0}$ as
the first index $r$ for which the bound $H_{r}(\Gamma)<h_{r}(N)$ is violated.
Then define $\ell$ as the first index such that
\[
|\mathrm{Int\,}(\gamma_{\ell})|\leq v_{r_{0}}(N)\,.
\]
To summarize, to any $\Gamma$ with length larger than $C\ln N$ we associate a
pair $(r_{0},\ell)$ and the subsequence of contours $\{\gamma_{\ell}%
,\gamma_{\ell+1},\dots,\gamma_{\ell+h_{r_{0}}-1}\}\subset\Gamma,$ for which
$\left(  \ref{61}\right)  $ holds. We denote by $\left(  r_{0},\ell
,\{\gamma_{\ell},\dots,\gamma_{\ell+h_{r_{0}}-1}\}\right)  $ the class of all
such hairs $\Gamma.$

The strategy of the proof will be to apply a Peierls type estimate (under the
volume constraint) to the section of the hair made of the exterior contours
$\{\gamma_{\ell},\dots,\gamma_{\ell+h_{r_{0}}-1}\}$.%

\begin{align}
& P_{\beta,N}\left\{  \mathcal{A}_{C}\right\}  \leq\label{62}\\
& \leq\sum_{r_{0}=1}^{R_{N}}\sum_{\ell=1}^{C_{3}\ln N}\;\sum_{\{\gamma_{i}%
\}}P_{\beta,N}\left\{  \varphi\text{ has a hair }\Gamma\text{ in the class
}\left(  r_{0},\ell,\{\gamma_{\ell},\dots,\gamma_{\ell+h_{r_{0}}-1}\}\right)
\right\}  \,,\nonumber
\end{align}
where the sum is over the collections of contours $\{\gamma_{\ell}%
,\dots,\gamma_{\ell+h_{r_{0}}-1}\}$. In order to estimate $\left(
\ref{62}\right)  $, two cases have to be distinguished. Either $r_{0}$ is
smaller than $R_{N}^{\prime}$ and all the contours $\gamma_{\ell},\dots
,\gamma_{\ell+h_{r_{0}}-1}$ are large, or $r_{0}$ is larger than
$R_{N}^{\prime}$ and one has to rely on more delicate estimates, taking into
account the fact that these contours might be small.

\vskip.5cm

\noindent Case 1 : $r_{0}\leq R_{N}^{\prime}$.\newline

In this case $h_{r_{0}}=4$ and we have%
\[
\sum_{i=0}^{3}|\mathrm{Int\,}(\gamma_{\ell+i})|\leq4v_{r_{0}}(N)\,,
\]%
\[
\sum_{i=0}^{3}|\gamma_{\ell+i}|\geq4(4\sqrt{v_{r_{0}+1}(N)})\,.
\]

Applying the Peierls estimate as in the proof of Theorem 6 (see relation
$\left(  \ref{44}\right)  $), we get for every $r_{0}=[0,R_{N}^{\prime}]$ and
every height $\ell\leq C_{3}\ln N$
\begin{align}
& \sum_{\{\gamma_{i}\}}P_{\beta,N}\left\{  \Gamma\in\left(  r_{0}%
,\ell,\{\gamma_{\ell},\dots,\gamma_{\ell+3}\}\right)  \right\}
\label{eq: partie 1}\\
& \qquad\leq\sum_{\{\gamma_{i}\}}\exp\left(  -\beta\sum_{i=0}^{3}|\gamma
_{\ell+i}|\right)  \ \frac{Q_{\beta,N}\left(  \mathbb{V}_{N}(\varphi
)\geq\lambda N^{3}-4v_{r_{0}}(N)\right)  }{Q_{\beta,N}\left(  \mathbb{V}%
_{N}(\varphi)\geq\lambda N^{3}\right)  }\,.\nonumber
\end{align}

The loss of volume can be compensated by adding a single square contour of
side length $4[\sqrt{4v_{r_{0}}(N)}]+1,$ in the same manner as in relation
$\left(  \ref{63}\right)  $ above. Furthermore the entropy of the four large
contours can be easily bounded, so that we obtain
\begin{align}
& \sum_{r_{0}=0}^{R_{N}^{\prime}}\sum_{\ell=0}^{C_{3}\ln N}\;\sum
_{\{\gamma_{i}\}}P_{\beta,N}\left\{  \Gamma\sim\left(  r_{0},\ell
,\{\gamma_{\ell},\dots,\gamma_{\ell+3}\}\right)  \right\}
\label{eq: hair large 1}\\
& \qquad\leq\sum_{r_{0}=0}^{R_{N}^{\prime}}\sum_{\ell=0}^{C_{3}\ln N}%
\;N^{8}\exp\left(  -16(\beta-\ln3)\sqrt{v_{r_{0}+1}(N)}+4\beta\sqrt{4v_{r_{0}%
}(N)}\right) \nonumber\\
& \qquad\leq\exp\left(  -C_{\beta}\ln N\right)  \,,\nonumber
\end{align}
where $C_{\beta}$ is a positive constant for $\beta$ large enough.

\vskip.5cm

\noindent Case 2 : $r_{0}>R_{N}^{\prime}$.\newline

First notice that for any collection of contours in the r.h.s. of $\left(
\ref{62}\right)  $, we have for the area
\[
\sum_{i=\ell}^{\ell+h_{r_{0}}-1}|\text{Int}(\gamma_{i})|\leq v_{r_{0}%
}(N)\,h_{r_{0}}(N)\,,
\]
while for the boundary we get
\[
\sum_{i=\ell}^{\ell+h_{r_{0}}-1}|\gamma_{i}|\geq4\sqrt{v_{r_{0}+1}(N)}%
h_{r_{0}}(N)\,.
\]

Since
\[
v_{r}(N)h_{r}(N)=a\frac{N^{2}}{2^{r}}C_{2}\,2^{r/2}N^{-1}\ln N=\frac{aC_{2}%
\,}{2^{r/2}}N\ln N,
\]
for $r>R_{N}^{\prime}=\frac{1}{\ln2}\ln\left(  \frac{aN^{2}}{C_{1}(\ln N)^{2}%
}\right)  $ we have
\[
v_{r}(N)h_{r}(N)<\frac{aC_{2}\,}{\sqrt{\frac{aN^{2}}{C_{1}(\ln N)^{2}}}}N\ln
N=\sqrt{aC_{1}}C_{2}(\ln N)^{2}.
\]
On the other hand,
\begin{equation}
\sqrt{v_{r+1}(N)}h_{r}(N)=\sqrt{a\frac{N^{2}}{2^{r+1}}}C_{2}\,2^{r/2}N^{-1}\ln
N=\sqrt{\frac{a}{2}}C_{2}\,\ln N.\label{65}%
\end{equation}
Thus the energy $\sum_{i=\ell}^{\ell+h_{r_{0}}-1}|\gamma_{i}|$ of the
collection of contours exceeds
\[
4\sqrt{\frac{a}{2}}C_{2}\,\ln N,
\]
and therefore, due to $\left(  \ref{64}\right)  ,$ is much larger than the one
of a single square contour with the same area, since its energy equals
\[
4\sqrt{v_{r_{0}}(N)h_{r_{0}}(N)}<4\sqrt{\sqrt{aC_{1}}C_{2}}\ln N.
\]
Nevertheless these contours can be small, and one has to estimate their
entropy carefully.

The number $\mathcal{N}(L_{\ell},\dots,L_{\ell+h_{r_{0}}-1})$ of compatible
contours $\{\gamma_{\ell},\dots,\gamma_{\ell+h_{r_{0}}-1}\}$ with respective
length $\{L_{\ell},\dots,L_{\ell+h_{r_{0}}-1}\}$ can be estimated by
\[
\mathcal{N}(L_{\ell},\dots,L_{\ell+h_{r_{0}}-1})\leq N^{2}\,3^{L_{\ell}%
}\,\prod_{i=\ell+1}^{\ell+h_{r_{0}}-1}\left(  v_{r_{0}}(N)\,3^{L_{i}}\right)
\,.
\]
To see this, one chooses first the contour $\gamma_{\ell}$ of length $L_{\ell
}$ at a random position in the box, then the other contours pile up above it,
so their entropy is given by the second term, where $v_{r_{0}}(N)$ is the
maximal area of $\gamma_{\ell}$.

We proceed as in $\eqref{eq: partie 1}$ and get for a given $r_{0}$ and height
$\ell$
\begin{align*}
& \sum_{\{\gamma_{i}\}}P_{\beta,N}\left\{  \Gamma\in\left(  r_{0}%
,\ell,\{\gamma_{\ell},\dots,\gamma_{\ell+h_{r_{0}}-1}\}\right)  \right\} \\
& \leq\sum_{\{\gamma_{i}\}}\exp\left(  -\beta\sum_{i=0}^{h_{r_{0}}-1}%
|\gamma_{\ell+i}|\right)  \ \frac{Q_{\beta,N}\left(  \mathbb{V}_{N}%
(\varphi)\geq\lambda N^{3}-v_{r_{0}}(N)h_{r_{0}}(N)\right)  }{Q_{\beta
,N}\left(  \mathbb{V}_{N}(\varphi)\geq\lambda N^{3}\right)  }\\
& \leq\sum_{\{L_{i}\}}\,N^{2}\exp\left(  h_{r_{0}}(N)\ln\left(  v_{r_{0}%
}(N)\right)  +(\ln3-\beta)\sum_{i=\ell}^{\ell+h_{r_{0}}-1}L_{i}+4\beta
\sqrt{v_{r_{0}}(N)h_{r_{0}}(N)}\right)  \,,
\end{align*}
where each $L_{i}$ ranges in $[4\sqrt{v_{r_{0}+1}(N)},4\sqrt{v_{r_{0}}(N)}],$
and so
\[
\sum_{i=\ell}^{\ell+h_{r_{0}}-1}L_{i}\geq4\sqrt{v_{r_{0}+1}(N)}\,h_{r_{0}}(N).
\]

Summing over the sequence $\{L_{i}\}$, we see that
\begin{align}
& \sum_{\{\gamma_{i}\}}P_{\beta,N}\left\{  \Gamma\sim\left(  r_{0}%
,\ell,\{\gamma_{\ell},\dots,\gamma_{\ell+h_{r_{0}}-1}\}\right)  \right\}
\label{eq: partie 2}\\
& \leq N^{2}\exp\left(  h_{r_{0}}(N)\sqrt{v_{r_{0}+1}(N)}\left(  \frac
{\ln\left(  v_{r_{0}}(N)\right)  }{\sqrt{v_{r_{0}+1}(N)}}-4(\beta
-10)+4\beta\sqrt{\frac{2}{h_{r_{0}}(N)}}\right)  \right)  \,.\nonumber
\end{align}

Recall that $h_{r_{0}}(N)>h_{R_{N}^{\prime}}(N)\geq10$ (see $\left(
\ref{64}\right)  $), $h_{r_{0}}(N)\sqrt{v_{r_{0}+1}(N)}=\sqrt{\frac{a}{2}%
}C_{2}\ln N$ (see $\left(  \ref{65}\right)  $), while $\frac{\ln\left(
v_{r_{0}}(N)\right)  }{\sqrt{v_{r_{0}+1}(N)}}\leq1$. Thus summing $\left(
\ref{eq: partie 2}\right)  $ over the indices $r_{0}$ and $\ell$, we obtain
\begin{align*}
& \sum_{r_{0}=R_{N}^{\prime}+1}^{R_{N}}\sum_{\ell=0}^{C_{3}\ln N}%
\;\sum_{\{\gamma_{i}\}}P_{\beta,N}\left\{  \Gamma\sim\left(  r_{0}%
,\ell,\{\gamma_{\ell},\dots,\gamma_{\ell+h_{r_{0}}-1}\}\right)  \right\} \\
& \leq\sum_{r_{0}=R_{N}^{\prime}+1}^{R_{N}}\sum_{\ell=0}^{C_{3}\ln N}N^{2}%
\exp\left(  \sqrt{\frac{a}{2}}C_{2}\ln N\left(  1-4(\beta-10)+\frac{4\beta
}{\sqrt{5}}\right)  \right) \\
& \leq\exp\left(  -C_{\beta}\ln N\right)  \,,
\end{align*}
where $C_{\beta}$ is positive constant for $\beta$ large enough.

Combining this with $\left(  \ref{eq: hair large 1}\right)  $, we conclude the proof.

\textbf{Acknowledgment. }\textit{We would like to thank P. Ferrari, D. Ioffe,
M. Pr\"{a}hofer and H. Spohn for very interesting discussions on step
fluctuations.\smallskip\ R.S. acknowledges the financial support of the NSF
grant DMS-0300672. S.S. acknowledges the financial support of the RFFI grant
03-01-00444.}


\begin{thebibliography}{9999}                                                                                             %
\bibitem[B]{B}T. Bodineau. \textit{The Wulff construction in three and more
dimensions}, Comm. Math. Phys. \textbf{207}, 197--229 (1999).

\bibitem[BFL]{BFL}J. Bricmont, J.-R. Fontaine, and J.L. Lebowitz:
\textit{Surface tension, percolation, and roughening}, J. Statist. Phys. 29,
no. 2, 193--203 (1982).

\bibitem[BFM]{BFM}J. Bricmont, J. Frohlich and A. El Mellouki: \textit{Random
surfaces in statistical mechanics: roughening, rounding, wetting, ... }, J.
Statist. Phys. \textbf{42}, no. 5-6, 743--798 (1986).

\bibitem[BI]{BI}T. Bodineau and D. Ioffe, \textit{\ Stability of interfaces
and stochastic dynamics in the regime of partial wetting}, preprint (2003).

\bibitem[BIV]{BIV}T. Bodineau, D. Ioffe and Y. Velenik: \textit{Rigorous
probabilistic analysis of equilibrium crystal shapes, }J. Math. Phys.
\textbf{41,} 1033--1098 (2000).

\bibitem[CP]{CP}R. Cerf and A. Pizstora. \textit{On the Wulff crystal in the
Ising model}, Ann. Probab. \textbf{28}, no. 3, 947--1017 (2000).

\bibitem[CK]{CK}R. Cerf and R. Kenyon: \textit{The low-temperature expansion
of the Wulff crystal in the 3D Ising model}, Comm. Math. Phys. \textbf{222},
no. 1, 147--179 (2001).

\bibitem[D]{D}R. L. Dobrushin: \textit{The Gibbs state that describes the
coexistence of phases for a three-dimensional Ising model}. (In Russian) Teor.
Verojatnost. i Primenen. \textbf{17}, 619--639 (1972).

\bibitem[DKS]{DKS}R.L. Dobrushin, R. Kotecky and S. B. Shlosman: \textit{Wulff
construction: a global shape from local interaction, }AMS translations series,
Providence (Rhode Island), 1992.

\bibitem[FPS]{FPS}P.~Ferrari, M.~Praehofer and H.~Spohn, \emph{Fluctuations of
an Atomic Ledge Bordering a Crystalline Facet}, preprint, (cond-mat/0303162)

\bibitem[FS]{FS}P.~Ferrari and H.~Spohn, \emph{Step fluctuations for a faceted
crystal}, preprint, (cond-mat/0212456).

\bibitem[FrSp]{FrSp}J. Fr\"{o}hlich and T. Spencer, \textit{\ The
Kosterlitz-Thouless transition in two-dimensional abelian spin systems and the
Coulomb gas}, Comm. Math. Phys. \textbf{81}, no. 4, 527--602 (1981).

\bibitem[IS]{IS}D. Ioffe and R. Schonmann,
\textit{Dobrushin-Koteck\'{y}-Shlosman theory up to the critical
temperature\/}, Comm. Math. Phys. \textbf{199}, 117--167 (1998).

\bibitem[MRSV]{MRSV}Ch. Maes, F. Redig, S. Shlosman and A. van Moffaert:
\textit{Percolation, Path Large Deviations and Weakly Gibbs States, }Comm.
Math. Phys., \textbf{209}, 517-545 (2000).

\bibitem[M]{M}S. Miracle-Sole. \textit{Surface tension, step free energy and
facets in the equilibrium crystal.} J.Stat. Phys., \textbf{79, }183-214 (1995).

\bibitem[SS1]{SS1}R.H. Schonmann and S. Shlosman: \textit{Constrained
variational problem with applications to the Ising model}, J. Stat. Phys.
\textbf{83}, 867--905 (1996).

\bibitem[SS2]{SS2}R.H. Schonmann and S. Shlosman:\textbf{\ }\textit{Complete
analyticity for 2D Ising completed.} Comm. Math. Phys. \textbf{170}, 453--482 (1995).

\bibitem[S1]{S1}S. Shlosman: \textit{The} \textit{Wulff construction in
statistical mechanics and in combinatorics, }arXiv.org e-Print archive,
math-ph/0010039, Russ. Math. Surv., \textbf{56}, no. 4, 709-738 (2001).

\bibitem[S2]{S2}S. Shlosman: \textit{Zero temperature Ising crystal, }in preparation.

\bibitem[VKer]{VKer}A. Vershik and S. Kerov. \textit{Asymptotic of the largest
and typical dimensions of irreducible representations of a symmetric group.}
Funct. Anal. Appl., \textbf{19}, 21-31 (1985).
\end{thebibliography}
\end{document}